# Influence of low-lying resonances on reaction rates for $n^{10}$B capture

S B Dubovichenko[1,2,*], N A Burkova[2], A V Dzhazairov-Kakhramanov[1,*]

[1] Fesenkov Astrophysical Institute "NCSRT" ASA MDASI RK, 050020, Almaty, RK
[2] al-Farabi Kazakh National University, RK, 050040, Almaty, RK
E-mail: dubovichenko@mail.ru, albert-j@yandex.ru

**Abstract:** The possibility of description available experimental data for total cross section for reaction of the neutron capture on $^{10}$B at energies from 25.3 meV to 1.0 MeV was considered in the frame of the modified potential cluster model with forbidden states, classification of the orbital cluster states according to Young diagrams and taking into account the resonance behavior of phase shifts of the $n^{10}$B elastic scattering. The obtained results agree with the experimental measurements. The calculated total cross sections are used for obtaining the reaction rate of the neutron capture on $^{10}$B at the temperature range from 0.01 to $10 \cdot T_9$. The reaction rate rises essentially at temperatures above 0.3–0.4 $T_9$, conditioned by the resonances at the energy above 100 keV.

## 1. Introduction

Necessity of study processes neutron radiative capture on boron isotopes $^{10,11}$B$(n,\gamma)$ lies at the interface of interests either fundamental, scientific (primordial nucleosynthesis of the Universe) [1–5] or practical, applied (exploitation of nuclear reactors) [6]. The reliable information on such reaction rates depending on temperature in the wide range required for further development of these directions. Synthesis of these interests is based on the fact that on the one hand isotopes $^{10,11}$B are stable in the ratio 20% for $^{10}$B and 80% for $^{11}$B, and are using as constructive elements for installations of different applications. On the other hand, these isotopes are stable components of the chain of some important astrophysical nuclear processes.

In the first case, specificity is connected with the estimation of fracture rate of protective reactor walls that reflect neutrons, which initially contain $^{11}$B in the capacity of *reflector* in view of littleness of absorption cross-section of thermal neutrons (0.05 barn). Isotope $^{10}$B mainly using as addition for *carbon absorptive* reactor rods, because rate of absorption of thermal neutrons is huge, about 3800 barn [6]. Anyway, estimations are needed for these reaction rates that lead to the elimination, i.e., replacement of the initial components of the constructive elements of reactors. In this case, the chain of displace reactions do not stop on $^{11}$B, because, it is possible that rates of synthesis heavier isotopes in the condition of compact neutron environments can be comparable.

The same circle of questions, although in somewhat other area, is a point at issue of the role of chain $^{10}$B$(n,\gamma)^{11}$B$(n,\gamma)^{12}$B$(n,\gamma)^{13}$B$(n,\gamma)^{14}$B$(n,\gamma)^{15}$B$(\beta^-)^{15}$C reactions with the efficiency to the carbocycle in stars (including evolution of supernovae at the stage of pre-bang etc.) [7–10]. According to modern independent assessments the disagreements of estimations of rates for considered reactions is equal to 100% [2]. In this context we note that known experimental projects, such us collaborations HELIOS [11] and FAIR [12] are aimed to studying of spectroscopic characteristics of short-half-life neutron-excess isotopes, including boron family.

---
[*] Corresponding authors: albert-j@yandex.ru, dubovichenko@mail.ru

Therefore, continuing studying processes of the radiative capture [13,14], we will consider the reaction $n+{}^{10}B \rightarrow {}^{11}B+\gamma$ at thermal and astrophysical energies in the frame of modified potential cluster model (MPCM) with forbidden states (FSs). Such potential two-cluster model uses conception of the forbidden states [15] and directly takes into account resonance behavior of the elastic scattering phase shifts of the considering particles at low energies [13,14]. The classification of the orbital cluster states according to Young diagrams, which allows one to find out the number of FSs and allowed states (ASs), i.e., the number of nodes in the wave function (WF) of the relative cluster motion. As usual, for bound states (BSs) or ground states (GSs) of $^{11}B$ that forming as a result of capture reaction in the cluster channel, which coincides with initial particles. The $n{}^{10}B$-potentials are constructed on the basis of description of the binding energy of these particles and some basic characteristics of such states, for example, charge radius and asymptotic constant (AC) [13,14].

New data on spectra of $^{11}B$ from [16] are used in the present calculations as against with our previous paper [17], where results of earlier review [18] were used. In this case, as against with our previous work, we take into account resonances of three scattering phase shifts of initial particles of the input channel. In other words, potentials of the $n{}^{10}B$ interactions for scattering processes describe now basic resonance states of the final nucleus $^{11}B$, which considering in the $n{}^{10}B$ channel at positive energy up to 1 MeV. Partial scattering potentials and BSs sometimes contain FS, constructed in the Gaussian form and obviously depend on all moments *JLS* of cluster system and orbital Young diagrams {*f*} of their relative motion [15]. Note that furthermore only BSs of $^{11}B$ of the negative parity are considered, which do not, apparently, halo-states [19].This, continually used by us approach, already justified itself at the considering more than 30 most different reactions of radiative capture type [13,14,20–27].

## 2. Classification of the $^{11}B$ states according to Young diagrams

Because there are no total product tables of Young diagrams for system with the number of particles more than eight [28], which are used by us earlier for similar calculations [13,14,20–24], so obtained further results should to consider as the qualitative estimation of possible orbital symmetries in the ground states $^{11}B$ for the considered here $n{}^{10}B$ channel. However, exactly on the basis of similar classification it is managed to quite acceptable explain available experimental data on the neutron radiative capture for the wide reaction circle [14,20–24]. Therefore, here we will use the classification of states on orbital symmetries, which leads to certain number of FSs and ASs in partial intercluster potentials, and consequently to certain number of nodes for wave function of relative motion.

It is possible to use Young orbital diagram in the form {442} for $^{10}B$, therefore for the $n{}^{10}B$ system we have {1} × {442} → {542} + {443} + {4421} [28]. The first from the obtained diagrams compatible with orbital moments *L* = 0, 2 etc. and is forbidden, because it cannot be five nucleons in the *s*-shell [15,29]. The second diagram is allowed and compatible with orbital moments $L = 1, 2$ etc. and the third, also, apparently, allowed compatible with $L = 1, 2, 3$ [29]. As it was said, the absence of product tables of Young diagrams for particles with number of 10 and 11 make it impossible gives the exact classification of cluster states in the considered system of particles. However, even so qualitative estimation of orbital symmetries allows one to determine the existence of forbidden state in $S$ and $D$ waves and absence of forbidden state for $P$ states. Exactly this structure of forbidden states and allowed states in different partial waves allows further one to construct potentials of intercluster interaction that are necessary for calculation of the total cross sections of the considered radiative capture reaction.

In such a way, limiting by lower partial waves with orbital moment $L = 0, 1$, we can say that for the $n^{10}$B system (for $^{10}$B is known $J^{\pi}, T = 3^{+}, 0$ [30]) only allowed state there is in potentials of the $P$ waves, and the forbidden state there is in $S$ waves. The state in the $^{6}P_{3/2}$ wave (in notations $^{(2S+1)}L_J$) corresponds to the GS of $^{11}$B with $J^{\pi}, T = 3/2^{-}, 1/2$ and lies at the binding energy of the $n^{10}$B system of -11.4541(2) MeV [16]. Some $n^{10}$B scattering states and BSs can be mixed by isospin with $S = 5/2$ ($2S+1 = 6$) и $S = 7/2$ ($2S+1 = 8$) that is taking into account further.

In addition, the scattering states in the $P$ wave depend up two Young diagrams {443} and {4421}, and the ground state, which also exists in the $P$ wave, apparently, from {443} only, as it was for some other cluster systems [14,15]. Therefore, $P$ state potentials of the $n^{10}$B scattering and BSs depend on the different number of Young diagrams, they can be also different for states with the same quantum numbers $JLS$.

Now let us consider excited states (ESs) available in $^{11}$B, but bound in the $n^{10}$B channel – these results are listed in Table 11.18 of work [16]:

1. The first excited, but bound in this channel state (first ES) with moment $J^{\pi} = 1/2^{-}$, which can be matched $F_{1/2}$ wave without FS there is at the energy of 2.124693(27) MeV above GS or -9.3294 MeV [16] relatively to the threshold of the $n^{10}$B channel. The $E1$ transitions are possible here from nonresonance $D_{1/2,3/2}$ scattering waves, but their influence is small and we do not consider this state.

2. The second ES at the energy of 4.44498(7) MeV [16] relatively to the GS or -7.0091 MeV relatively to the threshold of the $n^{10}$B channel has $J^{\pi} = 5/2^{-}$ and it can be matched to the mixture of $^{6}P_{5/2}$ and $^{8}P_{5/2}$ waves without FSs. Because the used model does not allow evidently divide states with different spin, furthermore the unified potential of such mixed state $^{6+8}P_{5/2}$ is constructed, i.e., the same potential is used for both spin states. Wave function, obtained with this potential by solving the Schrödinger equation and, in principle, consisted of two components for different spin channels, does not explicitly divide for these components in this model.

3. The third ES at the energy of 5.0203(3) MeV [16] relatively to the GS or -6.4338 MeV relatively to the threshold of the channel has $J^{\pi} = 3/2^{-}$ and it can be matched to the $^{6}P_{3/2}$ wave without FS.

4. The fourth ES at the energy of 6.74185(8) MeV [16] relatively to the GS or -4.7123 MeV relatively to the threshold of the channel has $J^{\pi} = 7/2^{-}$ and it can be matched to the mixture of $^{6}P_{7/2}$ and $^{8}P_{7/2}$ waves without FSs.

5. In addition, it is possible to consider ninth ES at the energy of 8.92047(11) MeV with the moment $5/2^{-}$, i.e., at the energy of -2.5336 MeV relatively to the $n^{10}$B threshold, which can be matched to the mixture of $^6P_{5/2}$ and $^8P_{5/2}$ states without FSs. There are experimental data [31] for the capture to this state, therefore we include it to the consideration.

Let us consider the spectrum of resonance states (RSs) in the $n^{10}B$ system, i.e., states at the positive energies (see Table 11.18 in work [16]):

1. The first RS of $^{11}B$ at the excitation energy of 11.600(20) MeV or 146(20) keV relatively to the threshold of the $n^{10}B$ channel has $J^\pi = 5/2^+$ at the total width of 180(20) keV and neutron width of 4 keV [16,18]. It can be matched to the $^6S_{5/2}$ scattering wave with FS or mixed by spin $^{6+8}D_{5/2}$ scattering wave also with the bound FS.

2. The second RS has the excitation energy of 11.893(13) MeV or 439(13) keV relatively to the threshold of the channel. Its total width of 194(6) keV, neutron width of 31 keV [18] and moment $J^\pi = 5/2^-$. Therefore, it can be matched to the mixtured $^{6+8}P_{5/2}$ scattering waves without FSs.

3. The third RS has the excitation energy of 12.040(130) MeV or 586(13) keV relatively to the threshold. Its total width is about of 1000 keV [16,18], neutron width changing from 770 keV [18] to 1.34 MeV [16], and moment is equal to $J^\pi = 7/2^+$ - it can be matched to the $^8S_{7/2}$ wave or $^{6+8}D_{7/2}$ scattering waves with FS.

4. The next resonance is located at the energy above 1 MeV, its moment does not fix unambiguously (see Table 11.18 in work [16]), and we do not consider it, because it is impossible to define its accessory to the certain partial wave.

The resonance levels that can be matched to the $P_{3/2}$ or $P_{7/2}$, $P_{9/2}$ states [16] are absence at the energy below 1 MeV in spectrum of $^{11}B$. Therefore, their phase shifts can be equalized to zero, and because there are no FSs in these waves, so the depth of potentials can be equalized to zero too [13,14]. Consequently, we will consider $E1$ transitions the GS and four (2$^{nd}$, 3$^{rd}$, 4$^{th}$ and 9$^{th}$) ES in $P$ waves from $S$ and $D$ scattering states and $M1$ process from the resonance $^6P_{5/2}$ scattering wave.

## 3. Methods of calculation of total cross sections

Total cross sections of the radiative capture $\sigma(NJ,J_f)$ for $EJ$ and $MJ$ transitions in potential cluster model are given, for example, in [32] or [13,14,33,34] and have the next form

$$\sigma(NJ,J_f) = \frac{8\pi Ke^2}{\hbar^2 q^3} \frac{\mu}{(2S_1+1)(2S_2+1)} \frac{J+1}{J[(2J+1)!!]^2} A_J^2(NJ,K) \sum_{L_i,J_i} P_J^2(NJ,J_f,J_i) I_J^2(J_f,J_i),$$

where $q$ is the wave number of particles of the input channel, $S_1$, $S_2$ are the particle spins in the input channel, $K$, $J$ are the wave number and the moment of γ-quanta in the output channel.

The value $P_J^2$ has the form [13,14,32] for electric orbital $EJ(L)$ transitions (at $S_i = S_f = S$)

$$P_J^2(EJ, J_f, J_i) = \delta_{S_i S_f} \left[(2J+1)(2L_i+1)(2J_i+1)(2J_f+1)\right](L_i 0 J 0 | L_f 0)^2 \begin{Bmatrix} L_i & S & J_i \\ J_f & J & L_f \end{Bmatrix}^2,$$

$$A_J(EJ, K) = K^J \mu^J \left(\frac{Z_1}{m_1^J} + (-1)^J \frac{Z_2}{m_2^J}\right), \qquad I_J(J_f, J_i) = \langle \chi_f | r^J | \chi_i \rangle$$

and its values for considered transitions are listed in Table 1. Here $L_f$, $L_i$, $J_f$, $J_i$ are orbital and total moments of particle of the input (*i*) and output (*f*) channels, $m_1$, $m_2$ masses of particles in input channel, $I_J$ is integral taken over wave functions of the initial $\chi_i$ and final $\chi_f$ state, as functions of relative cluster motion with intercluster distance *r*.

For consideration of the magnetic *M*1(*S*) transition, caused by the spin part of the magnetic operator, the next expression [13,14] is used – its values for considered transitions are given in Table 2

$$P_1^2(M1, J_f, J_i) = \delta_{S_i S_f} \delta_{L_i L_f} \left[S(S+1)(2S+1)(2J_i+1)(2J_f+1)\right] \begin{Bmatrix} S & L & J_i \\ J_f & 1 & S \end{Bmatrix}^2,$$

$$A_1(M1, K) = i \frac{\hbar K}{m_0 c} \sqrt{3} \left[\mu_1 \frac{m_2}{m} - \mu_2 \frac{m_1}{m}\right], \qquad I_J(J_f, J_i) = \langle \chi_f | r^{J-1} | \chi_i \rangle, \qquad J=1,$$

where *m* is the total mass of the nucleus, $\mu_1$ and $\mu_2$ are magnetic moments of clusters.

Let us consider possible transitions in the process of neutron radiative capture on $^{10}$B. Therefore, the GS and third ES of $^{11}$B the $^6P_{3/2}$ level is matched, so *E*1 transitions from the $^6S_{5/2}$ scattering wave to these states $\sigma_0 = \sigma(^6S_{5/2} \to {}^6P_{3/2}^0)$ and $\sigma_3 = \sigma(^6S_{5/2} \to {}^6P_{3/2}^3)$ are considered. In addition, the *E*1 transitions from the $^6S_{5/2}$ and $^8S_{7/2}$ scattering waves to the second ES and ninth ES of $^{11}$B, which are the mixture of two $^{6+8}P_{5/2}$ states, are taken into account.

$$^6S_{5/2} \to {}^6P_{5/2}$$
$$^8S_{7/2} \to {}^8P_{5/2}.$$

Therefore, here we have transition from different by spin initial *S* states to different parts of certain WF of this BS, the cross section of these transitions will summarize, i.e., $\sigma_2 = \sigma(^6S_{5/2} \to {}^6P_{5/2}^2) + \sigma(^8S_{7/2} \to {}^8P_{5/2}^2)$ and $\sigma_9 = \sigma(^6S_{5/2} \to {}^6P_{5/2}^9) + \sigma(^8S_{7/2} \to {}^8P_{5/2}^9)$ [14].

Another *E*1 transition is possible from the $^6S_{5/2}$ and $^8S_{7/2}$ scattering waves to the fourth ES of $^{11}$B at $J^\pi = 7/2^-$:

$$^6S_{5/2} \to {}^6P_{7/2}$$
$$^8S_{7/2} \to {}^8P_{7/2}.$$

The cross sections of these transitions also are summarized, as in the previous case $\sigma_4 = \sigma(^6S_{5/2} \to {}^6P_{7/2}^4) + \sigma(^8S_{7/2} \to {}^8P_{7/2}^4)$. Furthermore, the values of algebraic coefficients $P^2$ [13,14] for the calculation of all considered here $E1$ transitions are given in Table 1. The form of these coefficients is defined in the expressions given further.

Table 1. Coefficients $P^2$ for $E1$ transitions to the GS and ESs. The transitions from the resonance states are marked by the bold font.

| No. | $[^{(2S+1)}L_J]_i$ | Type of the transition | $[^{(2S+1)}L_J]_f$ | $P^2$ |
|---|---|---|---|---|
| 1. | $^6S_{5/2}$ | $E1$ | $^6P_{3/2}$, GS and third ES | 4 |
| 2. | $^6D_{1/2}$ | $E1$ | $^6P_{3/2}$, GS and third ES | 12/5 |
| 3. | $^6D_{3/2}$ | $E1$ | $^6P_{3/2}$, GS and third ES | 84/25 |
| **4.** | $^6\mathbf{D_{5/2}}$ | $\mathbf{E1}$ | $^6\mathbf{P_{3/2}}$, **GS and third ES** | **56/25** |
| 5. | $^6S_{5/2}$ | $E1$ | $^6P_{5/2}$ second ES and ninth ES | 6 |
| 6. | $^6D_{3/2}$ | $E1$ | $^6P_{5/2}$ second ES and ninth ES | 36/25 |
| **7.** | $^6\mathbf{D_{5/2}}$ | $\mathbf{E1}$ | $^6\mathbf{P_{5/2}}$ **second ES and ninth ES** | **768/175** |
| **8.** | $^6\mathbf{D_{7/2}}$ | $\mathbf{E1}$ | $^6\mathbf{P_{5/2}}$ **second ES and ninth ES** | **216/35** |
| 9. | $^8S_{7/2}$ | $E1$ | $^8P_{5/2}$ second ES and ninth ES | 6 |
| 10. | $^8D_{3/2}$ | $E1$ | $^8P_{5/2}$ second ES and ninth ES | 24/5 |
| **11.** | $^8\mathbf{D_{5/2}}$ | $\mathbf{E1}$ | $^8\mathbf{P_{5/2}}$ **second ES and ninth ES** | **162/35** |
| **12.** | $^8\mathbf{D_{7/2}}$ | $\mathbf{E1}$ | $^8\mathbf{P_{5/2}}$ **second ES and ninth ES** | **18/7** |
| 13. | $^6S_{5/2}$ | $E1$ | $^6P_{7/2}$ fourth ES | 8 |
| **14.** | $^6\mathbf{D_{5/2}}$ | $\mathbf{E1}$ | $^6\mathbf{P_{7/2}}$ **fourth ES** | **4/7** |
| **15.** | $^6\mathbf{D_{7/2}}$ | $\mathbf{E1}$ | $^6\mathbf{P_{7/2}}$ **fourth ES** | **24/7** |
| 16. | $^6D_{9/2}$ | $E1$ | $^6P_{7/2}$ fourth ES | 12 |
| 17. | $^8S_{7/2}$ | $E1$ | $^8P_{7/2}$ fourth ES | 8 |
| **18.** | $^8\mathbf{D_{5/2}}$ | $\mathbf{E1}$ | $^8\mathbf{P_{7/2}}$ **fourth ES** | **18/7** |
| **19.** | $^8\mathbf{D_{7/2}}$ | $\mathbf{E1}$ | $^8\mathbf{P_{7/2}}$ **fourth ES** | **128/21** |
| 20. | $^8D_{9/2}$ | $E1$ | $^8P_{7/2}$ fourth ES | 22/3 |

Data for $M1$ transitions from the resonance $^{6+8}P_{5/2}$ waves to the GS and different ESs are given in Table 2. Only transitions from the resonance wave are taken into account here, and contributions of other nonresonance waves are considered to be small.

Table 2. Coefficients $P^2$ for $M1$ transitions to the GS and ESs. Only transitions from the resonance $^6P_{5/2}$ state are taken into account.

| No. | $[^{(2S+1)}L_J]_i$ | Type of the transition | $[^{(2S+1)}L_J]_f$ | $P^2$ |
|---|---|---|---|---|
| 1. | $^6P_{5/2}$ | $M1$ | $^6P_{3/2}$, GS and third ES | 28/5 |
| 2. | $^6P_{5/2}$ | $M1$ | $^6P_{5/2}$ second ES and ninth ES | 2883/70 |
| 3. | $^8P_{5/2}$ | $M1$ | $^8P_{5/2}$ second ES and ninth ES | 1215/14 |
| 4. | $^6P_{5/2}$ | $M1$ | $^6P_{7/2}$ fourth ES | 40/7 |
| 5. | $^8P_{5/2}$ | $M1$ | $^8P_{7/2}$ fourth ES | 54/7 |

In some cases there are transitions between the mixed by spin $^{6+8}D$ scattering states to the also mixed by spin $^{6+8}P$ GS and ES. Therefore, the record of total cross section in the form of simple sum of cross sections of the form

$$\sigma(E1) = \sigma(^6D_{3/2} \to\,^6P_{5/2}) + \sigma(^8D_{3/2} \to\,^8P_{5/2})$$

can be considered as simple doubling of the cross section. Therefore, the used model does not allow one to divide spin states evidently, so for the calculation of each part of such process the same WFs are used, mixed by the spin and obtained in the given potential. Consequently, only spin factors are different in such matrix elements. Really, there is only one transition from the scattering state to the BS of nucleus, rather than two different $E1$ processes. Therefore, the cross section can be presented in the form of averaging by transitions from the mixed by spin $D$ scattering waves to the mixed $P$ excited states of $^{11}$B in the $n^{10}$B channel. This approach, suggested by us earlier at the consideration in the neutron radiative capture on $^{14}$N [14], was used for some other reactions, and allows one to obtain good results for description of total cross sections for these reactions [13,14,20–24].

Therefore, in the same way as before [13,14,20–24], the total cross section of the capture process to the ES for electromagnetic $E1$ transitions we represent in the form

$$\sigma(E1) = [\sigma(^6D_{3/2} \to\,^6P_{5/2}) + \sigma(^8D_{3/2} \to\,^8P_{5/2})]/2 + [\sigma(^6D_{5/2} \to\,^6P_{5/2}) + \sigma(^8D_{5/2} \to\,^8P_{5/2})]/2 +$$
$$+ [\sigma(^6D_{7/2} \to\,^6P_{5/2}) + \sigma(^8D_{7/2} \to\,^8P_{5/2})]/2 + \sigma(^6S_{5/2} \to\,^6P_{5/2}) + \sigma(^8S_{7/2} \to\,^8P_{5/2})$$

for second and ninth ESs and

$$\sigma(E1) = [\sigma(^6D_{5/2} \to\,^6P_{7/2}) + \sigma(^8D_{5/2} \to\,^8P_{7/2})]/2 + [\sigma(^6D_{7/2} \to\,^6P_{7/2}) + \sigma(^8D_{7/2} \to\,^8P_{7/2})]/2 +$$
$$+ [\sigma(^6D_{9/2} \to\,^6P_{7/2}) + \sigma(^8D_{9/2} \to\,^8P_{7/2})]/2 + \sigma(^6S_{5/2} \to\,^6P_{7/2}) + \sigma(^8S_{7/2} \to\,^8P_{7/2})$$

for fourth ES. The averaging by transitions with the same total moment, but with different channel spin, was carried out here.

In the case of $E1$ transitions to the ground state and third ES, which are pure by spin states, the cross sections are written in the form

$$\sigma(E1) = \sigma(^6S_{5/2} \to\,^6P_{3/2}) + \sigma(^6D_{1/2} \to\,^6P_{3/2}) + \sigma(^6D_{3/2} \to\,^6P_{3/2}) + \sigma(^6D_{5/2} \to\,^6P_{3/2}).$$

There is no any averaging, because the cross sections summarize between states unmixed by spin.

The methods of construction intercluster partial potentials used here, with the given moments $JLS\{f\}$, particularly represented in [13,14,20–24,35]. The next values of particle masses $m_n$ = 1.008664916 [36] and $m(^{10}B)$ = 10.012936 amu [37] are used in the calculations, and constant $\hbar^2/m_0$ is equal to 41.4686 MeV·fm$^2$. Coulomb parameter $\eta = \mu Z_1 Z_2 e^2/(q\hbar^2)$ is presented in the form $\eta = 3.44476 \cdot 10^{-2} \cdot Z_1 Z_2 \mu/q$, where $q$ is the wave number represented in fm$^{-1}$ and determined by the energy of interacting particles in the input channel. The Coulomb potential at $R_{coul} = 0$ was written in the form $V_{coul}$(MeV) = 1.439975·$Z_1 Z_2/r$, where $r$ is the relative distance between particles of the input channel in fm. Magnetic moments $\mu_n$ = -1.913$\mu_0$ [36] и $\mu(^{10}B)$ = 1.800$\mu_0$ [18], where $\mu_0$ is the nuclear magneton. The value of 2.4277(499) fm [38] is known for the root-mean-square radius of $^{10}$B, and the value of 2.406(294) fm is given for $^{11}$B in the base [39].

## 4. Potentials of the $n^{10}$B interaction

For all partial $n^{10}$B interaction potentials, i.e., for each partial wave with the given set of numbers $JLS\{f\}$, the Gaussian form was used

$$V(JLS\{f\},r) = -V_{0,JLS\{f\}}\exp(-\alpha_{JLS\{f\}}r^2).$$

Note at once that we did not succeed to find potentials pointed above first and third resonances, if to allow their existence in the $S$ scattering waves. Therefore, we will consider existence, for example, first resonance in the $^{6+8}D_{5/2}$ wave, and this gives parameters listed in Table 3 for No. 1. For the third resonance in the scattering wave $^{6+8}D_{7/2}$ parameters are given in Table 3 for No. 3, and for its neutron width the average value ~1000 keV from [16] and [18] is taken.

Table 3. Parameters of the scattering potentials and BSs. Here $E_{res.}$ – energy of resonance, $E_b$ – binding energy, $<r^2>^{1/2}$ – root-mean-square radius of nucleus, $\Gamma_{cm}$ – resonance width, $C_w$ – dimensionless asymptotic constant of the BS.

| No. | $^{(2S+1)}L_J$ | $V_0$, MeV | $\alpha$, fm$^{-2}$ | $E_{res.}$ (cm), MeV | $\Gamma_{cm}$, keV | $E_b$, MeV | $C_w$ | $<r^2>^{1/2}$, fm | Scattering phase shift $\delta°$ at the resonance energy or at 1 MeV |
|---|---|---|---|---|---|---|---|---|---|
| 1 | $^{6+8}D_{5/2}$ Res. | 129.585 | 0.1 | 146 | 3.8 | --- | --- | --- | 90(1) |
| 2 | $^{6+8}P_{5/2}$ Res. | 3555.923 | 13.0 | 440 | 33 | --- | --- | --- | 90(1) |
| 3 | $^{6+8}D_{7/2}$ Res. | 23.278 | 0.0205 | 590 | 1010 | --- | --- | --- | 90(1) |
| 4 | $^6S$, $^8S$, no res. | 160.5 | 0.5 | --- | --- | --- | --- | --- | 0(1) |
| 5 | $^3P_{3/2}$ GS | 165.3387295 | 0.45 | --- | --- | -11.454100(1) | 1.53(1) | 2.44 | 178(1) |
| 6 | $^3P_{3/2}$ GS | 602.548378 | 2.0 | --- | --- | -11.454100(1) | 0.71(1) | 2.43 | 179(1) |
| 7 | $^{6+8}P_{5/2}$ second ES | 108.374127 | 0.3 | --- | --- | -7.009100(1) | 1.44(1) | 2.44 | 176(1) |
| 8 | $^6P_{3/2}$ third ES | 149.701247 | 0.45 | --- | --- | -6.433800(1) | 1.10(1) | 2.44 | 176(1) |
| 9 | $^{6+8}P_{7/2}$ fourth ES | 143.727488 | 0.45 | --- | --- | -4.712300(1) | 0.94(1) | 2.44 | 176(1) |
| 10 | $^{6+8}P_{5/2}$ ninth ES | 135.3949665 | 0.45 | --- | --- | -2.533600(1) | 0.70(1) | 2.44 | 174(1) |

The expression $\Gamma = 2(d\delta/dE)^{-1}$ was used for calculation of the resonance widths $\Gamma$ for scattering phase shift $\delta$. Furthermore, all parameters of scattering potentials and BSs, which were used for carrying out calculation of total capture cross section, are given in Table 3. The elastic scattering phase shifts of potentials No.1–No.3 are shown in Fig. 1. We are using parameters, obtained in [17] and listed as No.4 in Table 3, for potentials of the nonresonance $S$ scattering waves. They lead to zero scattering phase shifts at energies up to 1 MeV. Such potential should to contain bound FS, therefore it cannot have zero value of depth, however it leads to zero scattering phase shifts. Potentials of $D$ waves also have FSs, and in all $P$ waves from Table 3 FSs are absent.

For pure by spin potential of the GS of $^{11}$B in the $n^{10}$B channel we will accept parameters

No.5 from Table 3. The dimensionless constant $C_w = 1.53(1)$ at the interval 3–10 fm, charge radius of 2.44 fm and mass radius of 2.39 fm at the binding energy of $E_b = -11.454100$ MeV with an accuracy of the finite-difference method of $10^{-6}$ MeV [40]. The AC error is coming due to its averaging according to the pointed distance range. The scattering phase shift for such potential smoothly decreases from 180° to 178° at the energy changing from zero to 1.0 MeV.

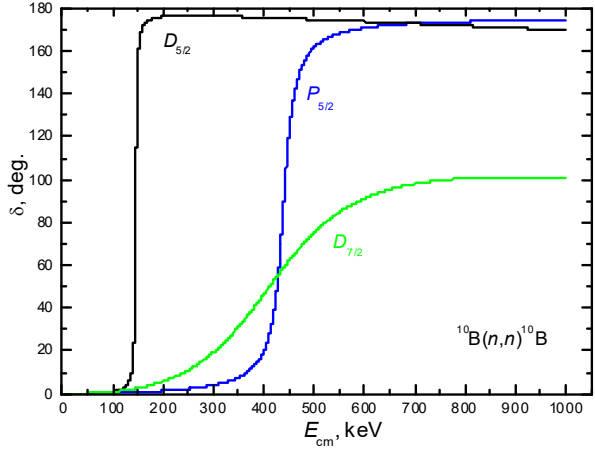

Fig. 1. Phase shifts of the $n^{10}B$ elastic scattering with resonances in $P$ and $D$ waves.

For the AC of the GS of $^{11}B$ in the cluster $n^{10}B$ channel in work [41] the value 1.72 fm$^{-1/2}$ was obtained. Somewhat other definition of the asymptotic constant was used in this work, namely $\chi_L(r) = C \cdot W_{-\eta L+1/2}(2k_0 r)$, which differ from our $\chi_L(r) = \sqrt{2k_0} \, C_w \cdot W_{-\eta L+1/2}(2k_0 r)$ [13,14] by the value $\sqrt{2k_0}$, equals 1.19 in this case, therefore for AC value in dimensionless value we have 1.44. In subsequent results for this AC [42] the specified value equals 1.82(15) fm$^{-1/2}$ is given, which after recalculation gives 1.52(12) in dimensionless form and completely matches with the value obtained here.

However, the spectroscopic factor $S_f$ in the $n^{10}B$ channel does not take into account in such estimations [17]. Its values in the interval from 0.81 to 5.05, with the average of the range 2.93(2.12) are given in review work [43]. Therefore, the asymptotic normalization coefficient $A_{NC}$ (ANC) connected with the dimension asymptotic constant $C$ in the following way $A_{NC} = \sqrt{S_f} C$, so taking for $A_{NC}$ the value of 1.82(15) fm$^{-1/2}$ and the average $\sqrt{S_f} = 1.57(67)$, we will obtain $C_w = 1.46(73)$ that is well agree with the value for potential No.5. However the other parameters are possible too, which, as it was seen further, describe well the available experimental data that given as No. 6 in Table 3. Bring to notice that the AC value of this potential is in the low limit of its possible values, given above.

The values given in Table 3 as No.7–No.10 are obtained for the parameters of all other potentials without FSs of four considered ESs of $^{11}B$ in the $n^{10}B$ channel. Note that parameters of two last potentials slightly changed relatively to results of [17], therefore new data for binding energy of these states come into view [16], in comparison with [18].

## 5. Total cross sections of the neutron radiative capture on $^{10}B$

In the first place the $E1$ transitions from $S$, $P$, and $D$ scattering waves with potentials No.1–No.4 to the GS with potential No.6 and to the ES with potentials No.7–No.10 from Table 3 were considered, and the obtained capture cross section is shown in Figs. 2a–2e by the solid curves. The total capture cross sections to the GS and different ESs from [31] at 23, 40 and 61 keV are shown by different points, and data of works [3,8,44] at 25.3 meV (1 meV = 10$^{-3}$ eV) are shown by triangles. The contribution of the $E1$ resonance from the first resonance at 146 keV is shown by the dotted curves in Fig. 2, and the contribution of the second resonance at 440 keV for $M1$ transitions is shown by the dashed curves. Exhibition of the third resonance at 590 keV is very small due to its large width of 1010 keV and can be seen only in Fig. 2b at the transition to the second ES with potential No.7, which is differ by width from thee following ES with potentials No.8–No.10.

At first, the potential No.5 was obtained for the GS, which leads to the $C_w$ that lies in the middle of the interval of possible AC values. However, results for the total cross sections of the transition to the GS with potential No.5 turn out to be much higher of the existence experimental data, therefore its parameters were changed for better description of the experiment – consequently, the potential No.6 was obtained. Potentials No.8–No.10 were obtained by simple decrease the depth of the GS potential No.5, therefore it does not succeed to find AC values for them. By the way, these potentials are able to correctly describe the value of total capture cross sections for these ESs. However, the same method led to the second ES potential, which did not describe available experimental results [31] at 23–61 keV and its parameters were changed for better description of these data – in such a way, the potential No.7 was obtained.

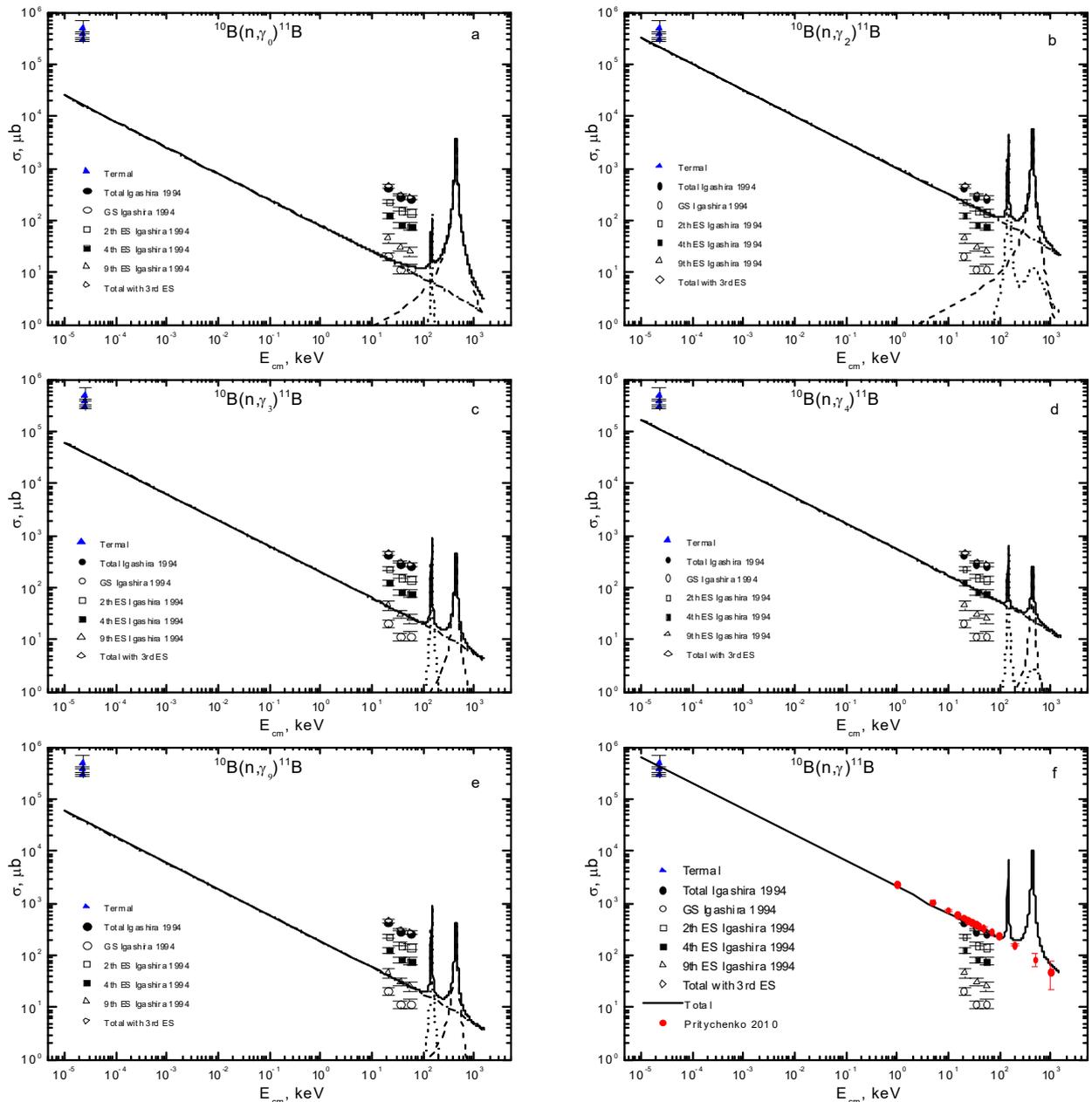

Fig. 2a,b,c,d,e,f. Total cross section of the neutron radiative capture reaction on $^{10}$B to the GS of $^{11}$B. Experimental data: black triangle (▲) – from works [3,8,44] at 25,3 meV, points (●) – total summarized cross sections of the neutron capture on $^{10}$B from [31], circles (o) – total capture cross sections to the GS, open squares (□) – total capture cross sections to the second ES, black squares (■) – total capture cross sections to the fourth ES and open

rhombs (◊) – summarized capture cross section from [31] taking into account transition to the third ES. Curves are described in the text. The cross sections from work [45] are shown by the red points.

All transitions listed in Tables 1 and 2 are taken into account in calculations, but the real contribution gives only transitions from resonance scattering states and the $E1$ transition from the nonresonance $S$ waves, contribution of which is shown in Fig. 2a–Fig. 2e by the dotted-dashed curves. Let us note that the transition to the third ES was not measured in work [31]. Our calculation lead to appropriate the same cross sections of the transitions to the third ES and ninth ES, as it was shown in Fig. 2c and Fig. 2e. Apparently, for obtaining more correct summarized total cross sections it is necessary to add cross sections to the ninth ES to the total cross sections from [31] once again, which is equivalent of taking into account transition to the third ES. Such cross sections are shown in all Figs. 2 by open rhombs – they, possibly, better coincide with results of our calculations, taking into account transition to the third ES.

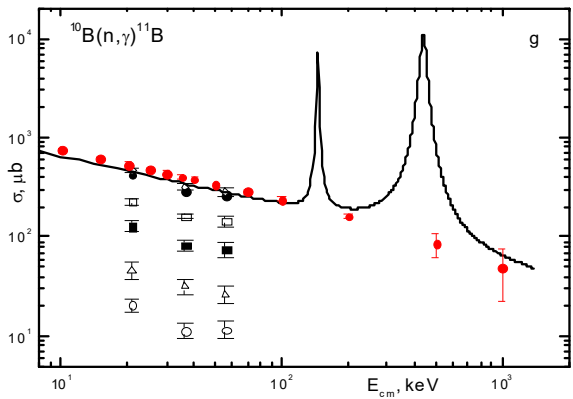

Fig. 2g. Total cross section of the neutron radiative capture reaction on $^{10}$B to the GS of $^{11}$B. Notations the same as in Fig. 2a – Fig. 2f.

The total summarized cross section taking into account all considered transitions is shown in Fig. 2f and Fig. 2g along with data [3,8,31,44]. The results of work [45], in which the comparison of total cross sections of certain reactions, are shown by the red points in Figs. 2f,g, including reaction considering here, obtained by different methods. These results do not take into account resonances, because of this the cross sections [45] at the energy more than 100 keV lie visibly lower than our calculations. The spread of results for different data from [45] is shown by the error band, and their middle value is given by point. As can be seen from Figs. 2f,g, the marked difference in total cross sections is seen only at the energies above 100 keV. The data from [45] at the energies below 100 keV, lie slightly above our results and experimental data of work [31].

Therefore, the calculated cross section at energies 10 meV to 10 keV in Fig. 2f is practically straight line; it can be approximated by the simple function of the form [13,14,20–24,26,27]

$$\sigma_{ap}(\mu b) = \frac{A}{\sqrt{E_{cm}(keV)}}.$$

The value of constant $A = 2037.58$ mkb·keV$^{1/2}$ was obtained in one point for calculated cross sections (the solid curve in Fig. 2f) at the minimal energy equals 10 meV. Module $M(E) = \left|[\sigma_{ap}(E) - \sigma_{theor}(E)]/\sigma_{theor}(E)\right|$ of the relative deviation of calculated theoretical cross section ($\sigma_{theor}$) and approximation ($\sigma_{ap}$) of this cross section by given above function in the range down to 10 keV is at the level of tenth of percent. It is quite real to suppose that this form of dependence total cross section from energy will save at lower energies too. Therefore, it is possible to do estimation of the cross section value, for example, at energy 1 mkeV (1 mkeV = $10^{-6}$ eV), which gives the cross section value about 65 b. From this expression, we obtain approximately 405 mb at new experimental value 394(15) mb [3] at thermal energy.

## 6. Reaction rate of the neutron radiative capture on $^{10}$B

If the total cross sections of certain reaction process are known, it is possible to determine the rate of this reaction, which usually uses in the consideration of astrophysical problems [32]

$$N_A \langle \sigma v \rangle = 3.7313 \cdot 10^4 \mu^{-1/2} T_9^{-3/2} \int_0^\infty \sigma(E) E \exp(-11.605 E / T_9) dE,$$

where $E$ is the energy in MeV, cross section $\sigma(E)$ in μb, μ is the reduced mass in amu, $T_9$ is the temperature in $10^9$ K.

On the basis of the obtained capture cross sections to all BSs, shown in Fig. 2, it is possible to calculate reaction rate, which corresponds to this capture – results are shown in Fig. 3a at the temperature range from 0.01 to 10 $T_9$. The cross sections in the range from 1 meV to 5 MeV, but without taking into account the resonances above 1 MeV, were used for calculation of rate. The summarized capture reaction rate to all BSs, which corresponds to Fig. 2f, is shown by the blue solid curve in Fig. 3a. Other curve show capture rates to all other considered above bound states.

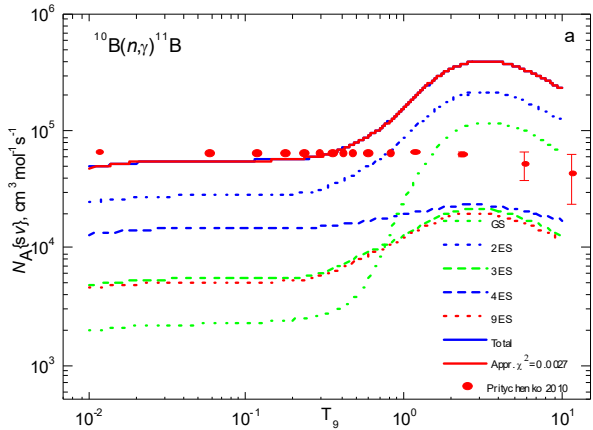 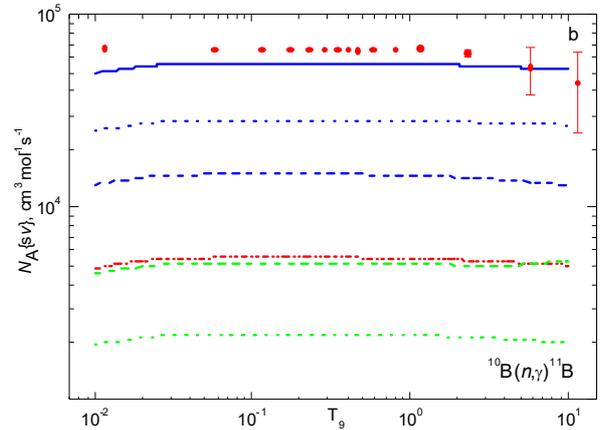

Fig. 3a. The reaction rate of the neutron radiative capture on $^{10}$B to all BSs of $^{11}$B. Curves – results of our calculations that are described in the text. The results of [45] are shown by the red points.

Fig. 3b. The reaction rate of the neutron radiative capture on $^{10}$B to all BSs of $^{11}$B. Curves – results of our calculations that do not take into account resonances. Notations the same as in the previous figure.

The results of work [45] are shown by the red points in Fig. 3a, where the estimation of rates for some reactions that are obtained on the basis of different data bases is done. The value range is shown by errors from different databases, and points show their middle value. It is seen that results of this work at the temperature more than 0.3–0.4 $T_9$ lead to reaction rate more lower than our results, because do not take into account existence resonances in total cross sections at energies above 100 keV, as it was shown by red points in Fig. 2f,g.

Furthermore, the parametrization of the obtained by us calculated cross section was done in the next form [46]

$$N_A \langle \sigma v \rangle = \frac{a_1}{T^{2/3}} \exp(-\frac{a_2}{T^{1/3}})(1 + a_3 T^{1/3} + a_4 T^{2/3} + a_5 T + a_6 T^{4/3} + a_7 T^{5/3} + a_8 T^{7/3} + a_9 T^{9/3}) +$$
$$+ \frac{a_{10}}{T^{1/2}} \exp(-\frac{a_{11}}{T^{1/2}}) + \frac{a_{12}}{T} \exp(-\frac{a_{13}}{T}) + \frac{a_{14}}{T^2} \exp(-\frac{a_{15}}{T^2})$$

Values of expansion parameters are listed in Table 4, value $T$ denotes $T_9$, and the error of predicted reaction rate for calculation of the $\chi^2$ is given at the level of 5%. Parameters for the reaction rate from Table 4 lead to the value of $\chi^2 = 0.0027$, and calculation results according this formula are shown in Fig. 3a by the red solid curve.

Furthermore, our results for reaction rates calculated without taking into account all three resonances are shown in Fig. 3b, as it is seen they lies slightly lower than data [45]. We considered only transitions to BSs of the negative parity for $^{11}$B, because exactly for them there are experimental data [31]. Therefore, the smaller values of total cross sections, obtained in our calculations, relatively to results of [45] can be, apparently, explain by the absence of accounting of transitions to the BSs of the positive parity. As it was seen before, total cross sections from work [45], shown in Figs. 2f,g, lie slightly above experimental data [31], therefore, the reaction rates have the greater values, than obtained by us in our calculations.

Table 4. Parameters of analytical parametrization of the reaction rate

| No. | $a_i$ |
|---|---|
| 1. | 0.9742840E-02 |
| 2. | 0.2737900E+01 |
| 3. | 0.2348600E+09 |
| 4. | -0.1639500E+09 |
| 5. | -0.3782000E+08 |
| 6. | 0.6766587E+03 |
| 7. | 0.8487100E+07 |
| 8. | 0.1585100E+07 |
| 9. | -0.2550300E+06 |
| 10. | 0.2486022E+05 |
| 11. | 0.1640631E+00 |
| 12. | 0.4631300E+07 |
| 13. | 0.3929600E+01 |
| 14. | 0.3173494E+05 |
| 15. | 0.5126532E+00 |
| $\chi^2 = 2.68$E-003 | |

## 7. Conclusion

As it is seen from the given results, the used methods allow one to obtain acceptable description of the available experimental data for total cross sections of the neutron capture on $^{10}$B from [3,8,31,44] at the energy range from 25.3 meV to 61 keV. That, only AC for the GS $^{11}$B in $n^{10}$B channel is known, the ACs for all ESs are absent, therefore, results of calculation results with transitions to 2, 3, 4 and 9-th ESs ought to consider preliminary. However, exactly presented above AC for these ESs can consider them as preliminary values, which were obtained in the present work at the consideration of the neutron radiative capture on $^{10}$B. In addition, it was found out that for the correct description of widths of the first and third resonances of the positive parity they have to be in the $D$ scattering wave.

The reaction rates at temperatures below 0.3–0.4 $T_9$ obtained by us, in whole, agree with the known earlier data from work [45]. Note, these results do not take into account resonances at energy above 100 keV, therefore the reaction rates of work [45] smoothly decrease at higher temperatures. According to our results at temperatures above 0.3–0.4 $T_9$ the visible rise of rates $\langle \sigma v \rangle$, caused by these resonances, is found. The obtained new shape of the reaction rate can essentially change the results of calculations of different macroscopic and astrophysical characteristics in processes of the neutron radiative capture on $^{10}$B.

Finally, it is necessary to note that available experimental data [3,31] evidently are not enough for the detail comparison with the calculation results in the resonance range. The new measurements of total cross sections at resonance energies are needed, i.e., approximately, from 100 and to 800 keV, with sufficient small step, so that it is possible to observe the shape of these resonances.

## Acknowledgments

This work was supported by the Grant of Ministry of Education and Science of the Republic of Kazakhstan through the program BR05236322 "Study reactions of thermonuclear processes in extragalactic and galactic objects and their subsystems" in the frame of theme "Study of thermonuclear processes in stars and primordial nucleosynthesis" through the Fesenkov Astrophysical Institute of the National Center for Space Research and Technology of the Ministry of Defence and Aerospace Industry of the Republic of Kazakhstan (RK).

## References

1. Igashira M and Ohsaki T 2004 *Sci. Tech. Adv. Materials* **5**, 567-573
2. Sasaqui T, Kajino T, Mathews G J, Otsuki K and Nakamura T 2005 *Astrophys. J.* **634**, 1173-1189
3. Firestone R B and Revay Zs 2016 *Phys. Rev. C* **93**, 054306
4. Lee H Y et al 2010 *Phys. Rev. C* **81**, 015802
5. Iwasaki H 2010 *Mod. Phys. Lett. A* **25**, 1967-1971
6. Baranov V Yu 2005 *Isotopes: properties, preparation, applications* (Moscow: Fizmatlit) 728p. (in Russian)
7. Wuosmaa H 2015 *Acta Phys. Pol. B* **46**, 627-637
8. Mughabghab S F 2006 *Atlas of neutron resonances* (Brookhaven: National Nuclear Data Center) 1008p.
9. Descouvemont P and Dufour M 2012 *Microscopic cluster model*, In: Clusters in Nuclei. V.2, Editor C. Beck, (Berlin-Heidelberg: Springer-Verlag) 353p.
10. Dubovichenko S B and Burkova N A 2014 *Mod. Phys. Lett. A* **29**, 1450036(1-14)
11. Back B B, Baker S I, Brown B A et al 2010 *Phys. Rev. Lett.* **104**, 132501(4p.).

12. Reifarth R, Altstadt1 S and Gobel K 2016 *Nuclear astrophysics with radioactive ions at FAIR, Nuclear Physics in Astrophysics. VI (NPA6)*, Jour. Phys.: Conf. Ser. **665**, 012044(12p.)
13. Dubovichenko S B 2015 *Thermonuclear processes in Stars and Universe*. Second English edition (Saarbrucken: Scholar's Press.) 332p.; https://www.scholars-press.com/catalog/details/store/gb/book/978-3-639-76478-9/Thermonuclear-processes-in-stars
14. Dubovichenko S.B. 2019 *Radiative neutron capture and primordial nucleosynthesis of the Universe*. First English edition. Germany. Berlin. De Gruyter 2019. 293p. ISBN 978-3-11-061784-9. https://doi.org/10.1515/9783110619607-201.
15. Neudatchin V G *et al* 1992 *Phys. Rev. C* **45**, 1512-1527
16. Kelley J H *et al* 2012 *Nucl. Phys. A* **880**, 88-195
17. Dubovichenko S B 2014 *Russ. Phys. J.* **57**, 880-887
18. Ajzenberg-Selove F 1990 *Nucl. Phys. A* **506**, 1
19. Demyanova A S *et al* 2016 *Jour. Phys.: Conf. Ser.* **724**, 012013
20. Dubovichenko S B and Dzhazairov-Kakhramanov A V 2012 *Int. Jour. Mod. Phys. E* **21**, 1250039(44p.)
21. Dubovichenko S B, Dzhazairov-Kakhramanov A V and Burkova N A 2013 *Int. Jour. Mod. Phys. E* **22**, 1350028(52p.)
22. Dubovichenko S B, Dzhazairov-Kakhramanov A V and Afanasyeva N V 2013 *Int. Jour. Mod. Phys. E* **22**, 1350075(53p.)
23. Dubovichenko S B and Dzhazairov-Kakhramanov A V 2014 *Int. Jour. Mod. Phys. E* **23**, 1430012(1-55)
24. Dubovichenko S B and Dzhazairov-Kakhramanov A V 2017 *Int. Jour. Mod. Phys. E* **26**, 1630009(56p.)
25. Dubovichenko S B and Dzhazairov-Kakhramanov A V 2016 *Astrophys. J.* **819**, 78(8p.)
26. Dubovichenko S B and Dzhazairov-Kakhramanov A V 2016 *J. Phys. G* **43**, 095201(14p.); 2017 *Corrigendum: J. Phys. G.* **44**, 079503
27. Dubovichenko S B, Dzhazairov-Kakhramanov A V and Afanasyeva N V 2017 *Nucl. Phys. A* **963**, 52-67
28. Itzykson C and Nauenberg M 1966 *Rev. Mod. Phys.* **38** 95
29. Neudatchin V G and Smirnov Yu F 1969 *Nucleon associations in light nuclei* (Moscow: Nauka) 414p. (in Russian)
30. Tilley D R *et al* 2004 *Nucl. Phys. A* **745**, 155-363
31. Igashira M *et al* 1994 *Measurements of kev-neutrons capture gamma rays*, Conf. Meas. Calc. and Eval. of Photon Prod. Data. Bologna, P. 269
32. Angulo C *et al* 1999 *Nucl. Phys. A* **656**, 3-183
33. Dubovichenko S B and Dzhazairov-Kakhramanov A V 1995 *Phys. Atom. Nucl.* **58**, 579-585
34. Dubovichenko S B and Dzhazairov-Kakhramanov A V 1997 *Phys. Part. Nucl.* **28**, 615-647
35. Dubovichenko S B and Dzhazairov-Kakhramanov A V 2015 *Nucl. Phys. A* **941**, 335-363
36. Neutron magnetic moment; https://physics.nist.gov/cgi-bin/cuu/Value?munn|search_for=atomnuc!

37. Nuclear Wallet Cards; http://cdfe.sinp.msu.ru/services/ground/NuclChart_release15.html
38. Chart of nucleus shape and size parameters, $^{10}$B (Z=5); http://cdfe.sinp.msu.ru/cgi-bin/muh/radcard.cgi?z=5&a=10&td=123456
39. Chart of nucleus shape and size parameters, $^{11}$B (Z=5); http://cdfe.sinp.msu.ru/cgi-bin/muh/radcard.cgi?z=5&a=11&td=123456
40. Dubovichenko S B 2012 *Methods of Calculation of Nuclear Characteristics: Nuclear and Thermonuclear Processes*. Second Russian Edition, corrected and enlarged (Saarbrucken: Lambert Acad. Publ. GmbH&Co. KG.) 432p.; https://www.lap-publishing.com/catalog/details//store/ru/book/978-3-659-21137-9/методы-расчета-ядерных-характеристик
41. Blokhintsev L D, Borbey I and Dolinsky E I 1977 *Phys. Part. Nucl.* **8**, 1189-1245
42. Yarmukhamedov R. 2013 *Determination of ANC for n$^{10}$B channel in $^{11}$B nucleus by $^{10}$B(d,p)$^{11}$B reaction at $E_d$ = 12 MeV*, Private Communication
43. Lee J, Tsang M B and Lynch W G 2007 *Phys. Rev. C* **75**, 064320(21p.)
44. Bartholomew G A, Campion P J 1957 *Can. J. Phys.* **35**, 1347
45. Pritychenko B, Mughaghab S F and Sonzogni A A 2010 *Atom. Data Nucl. Data Tab*. **96**, 645748
46. Caughlan G R and Fowler W A 1988 *Atom. Data Nucl. Data Tab*. **40**, 283-334